\documentclass[10pt,aps,twocolumn,nofootinbib,superscriptaddress]{revtex4-1}
\usepackage{amssymb}
\usepackage{amsmath}
\usepackage{latexsym}
\usepackage{verbatim}
\usepackage{graphicx}
\usepackage{epsfig}
\usepackage{stmaryrd}
\usepackage{rotating,graphicx}
\usepackage[english]{babel}
 \usepackage{setspace}
\usepackage[utf8]{inputenc}
\usepackage{natbib}
\usepackage[T1]{fontenc}
\usepackage{lmodern}
\usepackage{pstricks}

\newcommand{\ket}[1]{|#1\rangle}

\begin{document}

\title{Can gravity account for the emergence of classicality?}

\author{Yuri Bonder}
\email{bonder@nucleares.unam.mx}
\affiliation{Instituto de Ciencias Nucleares, Universidad Nacional Aut\'onoma de M\'exico\\
A. Postal 70-543, M\'exico D.F. 04510, M\'exico}

\author{Elias Okon}
\email{eokon@filosoficas.unam.mx}
\affiliation{Instituto de Investigaciones Filos\'oficas, Universidad Nacional Aut\'onoma de M\'exico\\
Circuito Maestro Mario de la Cueva s/n, M\'exico D.F. 04510, M\'exico}

\author{Daniel Sudarsky}
\email{sudarsky@nucleares.unam.mx}
\affiliation{Instituto de Ciencias Nucleares, Universidad Nacional Aut\'onoma de M\'exico\\
A. Postal 70-543, M\'exico D.F. 04510, M\'exico}

\begin{abstract}
A recent debate has ensued over the claim by Pikovski \textit{et al.} [Nat. Phys. \textbf{11}, 668 (2015)] that systems with internal degrees of freedom undergo a universal, gravity-induced, type of decoherence that explains their quantum-to-classical transition. This decoherence is supposed to arise from the different gravitational redshifts experienced by such systems when placed in a superposition of two wave packets at different heights in a gravitational field. Here we investigate some aspects of the discussion with the aid of simple examples. In particular, we first resolve an apparent conflict between the reported results and the equivalence principle by noting that the static and free-fall descriptions focus on states associated with different hypersurfaces. Next, we emphasize that predictions regarding the observability of interference become relevant only in the context of concrete experimental settings. As a result, we caution against hasty claims of universal validity. Finally, we dispute the claim that, at least in the scenarios discussed by Pikovski \textit{et al.}, gravitation is responsible for the reported results and we question the alleged ability of decoherence to explain the quantum-to-classical transition. In consequence, we argue against the extraordinary assertion by Pikovski \textit{et al.} that gravity can account for the emergence of classicality.
\end{abstract}

\maketitle

\section{Introduction}
\label{Int}

The interface between gravitation and quantum theory is a fascinating subject. It has inspired novel and exiting ideas, many of them adventuring beyond standard quantum mechanics or general relativity (see \textit{e.g.}, Refs. \onlinecite{Penrose1,Penrose2,Penrose3,Penrose4}). The topic is also riddled with subtleties and slight confusion can easily lead to questionable conclusions. Consider, in this regard, the seemingly paradoxical question of the emission of photons by a charged particle undergoing constant proper acceleration, as seen by an equally accelerated observer \cite{Bremstralung1,Bremstralung2,Bremstralung3}. In view of the equivalence principle, why would a static charge in a static gravitational field radiate? Other examples include the alleged violation of the equivalence principle by oscillating neutrinos \cite{neutrinos1,neutrinos2}, the fierce controversy surrounding the claim that the gravitational redshift can be measured with an atom interferometer \cite{nobels1,nobels2,nobels3}, or the proliferation of incompatible opinions regarding the validity of the equivalence principle in quantum mechanics \cite{EPandQM}.

Another dramatic case in the area, which has recently ignited a heated debate \cite{Adl.Bas:15,Bon.Oko.Sud:15,Dio:15,Pik:15b}, is provided by the article ``Universal decoherence due to gravitational time dilation'' \cite{Pik:15a}. There, it is claimed that gravity can produce a type of decoherence that renders invisible any interference between components of the state of a composite system located at different heights in a gravitational field. Based on that assertion, it is argued that gravitation is responsible for the quantum-to-classical transition of such systems. A noteworthy aspect of these claims is that, in contrast with other speculative ideas in this realm, they are supposed to follow directly from standard quantum theory and general relativity. For example, in Ref.~\cite[p. 4]{Pik:15a} the authors write: 
\begin{quotation}
Our results show that general relativity can account for the suppression of quantum behavior for macroscopic objects without introducing any modifications to quantum mechanics or to general relativity.
\end{quotation}
Needless to say, a result of these characteristics would be truly extraordinary. Unfortunately, as we argue below, there are reasons to doubt the validity of some of the claims in Ref.~\onlinecite{Pik:15a}.

It is important to stress, at the onset, that the validity of the Hamiltonian employed in Ref.~\onlinecite{Pik:15a} is not questioned in this manuscript; we take such a Hamiltonian for granted. What we are interested in is in the \emph{physical significance} of the results obtained and on a critical assessment of the validity of claims regarding the generality of such results, their relation to gravity and their ability to explain the emergence of classicality.

Part of the controversy surrounding Ref.~\onlinecite{Pik:15a} arises from an apparent conflict between the reported effect and the \emph{equivalence principle} \cite{Bon.Oko.Sud:15,Dio:15}. The point is that, according to such a principle, by analyzing the situation at hand using a free-falling frame, one deals with a situation with no gravity, and thus, with nothing to cause the alleged decoherence. However, the authors of Ref.~\onlinecite{Pik:15a} maintain that the effect is in fact frame and coordinate independent, and that, even in the absence of gravity, acceleration would lead to a similar result (see also Ref.~\onlinecite{Pik:15b}). The authors of Ref.~\onlinecite{Pik:15a} are of course correct in pointing out that the proper time between two events along a certain worldline is frame independent. Nevertheless, in Sec. \ref{exa}, using an extremely simple example, we argue that they fail to stress the dependence of the effect they study on the particular way in which one compares the proper times along two or more of such world-lines, and the rather arbitrary nature of the choices involved. As a result, we show that the apparent inconsistencies between the claimed effect and the equivalence principle are resolved by recognizing that the descriptions of the scenarios considered in Ref.~\onlinecite{Pik:15a}, as given on different frames, often correspond to different physical situations (also see Ref.~\onlinecite{Dio:15}).

Next, in order to further clarify the situation, in Sec. \ref{Wit}, we explore general conditions under which the reported effect appears. In this regard, as is correctly noted in Ref.~\onlinecite{Pik:15a}, we conclude that its essence lies in a difference of proper times between the initial and final events along the two worldlines of the superpositions considered. However, we press the point that, given one of the worldlines and the initial and final points on it, there is no canonical way to select the corresponding two points on the other worldline. Then, given the delicate dependence of the effect on this choice, we question the pertinence of most of the general results presented in Ref.~\onlinecite{Pik:15a}. More specifically, we argue that a crucial conceptual issue is systematically overlooked and that this drastically alters the physical significance of the reported results. This issue is based on the well-known fact that the comparison of quantum phases at different spacetime points is meaningless. The problem is that most, if not all, of the situations considered involve this kind of empty phase comparisons. As a result, we argue that the results are useful only in connection with concrete experimental settings, and we caution against hasty claims of generality and universal validity. 

Finally, by pointing out that spacetime curvature is never relevant in the scenarios considered in Ref.~\onlinecite{Pik:15a}, in Sec. \ref{sub} we argue against the assertion that gravitation is responsible for the cited effect. We also question the widespread belief that decoherence brings about the quantum-to-classical transition. In consequence, we challenge the far reaching assertion defended in Ref.~\onlinecite{Pik:15a} that gravity can account for the emergence of classicality. To wrap up, in Sec. \ref{Con} we present our conclusions.

\section{Time dilation, reference frames and the equivalence principle}
\label{exa}

In order to start the discussion, we explore a simple example that does not involve gravitation or acceleration. As we show below, the example is useful not only because it displays an effect analogous to those reported in Ref.~\onlinecite{Pik:15a} but also because it illuminates the issue of an apparent incompatibility between the reported effect and the equivalence principle. The idea is to consider the same type of system used in Ref.~\onlinecite{Pik:15a} consisting of a quantum particle with a relatively well-localized center of mass and with internal degrees of freedom characterized by the Hamiltonian $H_0$. As in Ref.~\onlinecite{Pik:15a}, the Hamiltonian (for one spatial dimension $x$) describing such a system, to a good approximation, is taken to be 
\begin{equation}\label{Hinitial}
 H= H _{\rm ext} + \left( 1+ \frac{\Phi (x)}{c^2} - \frac{p^2}{2m^2c^{2}}\right) H_0,
\end{equation}
where
\begin{equation}
 H_{\rm ext} = m c^2 + \frac{p^2}{2m} + m \Phi (x) .
\end{equation}
As it is customary, $m$, $p$, and $\Phi$, respectively, denote the mass, momentum, and Newtonian gravitational potential, while $c$ is the speed of light. Observe that the internal Hamiltonian $H_0$ has a correction due to both, the Newtonian gravitational potential and the system's momentum, and, in fact, the object multiplying $H_0$ can be recognized as the dominant (\textit{i.e.}, order $c^{-2}$) part of the factor linking proper and coordinate times. These terms must be kept even if one wants to consider the nonrelativistic limit because they are typically multiplied by the rest energy, which is of order $c^2$, and thus, they generate relevant terms. It is important to keep in mind that obtaining the nonrelativistic Hamiltonian for more complicated situations, like those involving unstable particles, is highly nontrivial \cite{unstable}. In any case, the key point is that, according to Eq.~(\ref{Hinitial}), velocity and gravity generate similar effects on the part of the evolution described by $H_0$. 

Consider the simple case where $\Phi =0$ and suppose that, at time $t=0$, according to the notion of time associated with the reference frame $K$, the system is described by
\begin{equation}
\label{InSt}
\ket {\Psi (0)} = \frac{1}{\sqrt{2}}\left[\chi_{p_0}(0)+\chi_{-p_0}( 0)\right]\otimes \ket{\psi(0)},
\end{equation}
where $\chi_p(x)$ represents a wave packet centered about $x$ with momentum $p$ and $\ket{\psi(0)}$ is an arbitrary state for the internal degrees of freedom given by $\ket{\psi(0)}=\sum_i \alpha_i \ket{\psi_i}$ with $H_0\ket{\psi_i}=E_i \ket{\psi_i}$. The initial state (\ref{InSt}) is a superposition of two wave packets centered about $x =0$, one with momentum $p_0$ and the other with momentum $-p_0$, both with the same internal state (see Fig.~\ref{fig1}). Note that this state is separable. It is clear that, after a time $T$, the state (omitting a global phase) is given by 
\begin{equation}
\ket {\Psi (T)} = \frac{1}{\sqrt{2}}\left[\chi_{p_0}\left(\frac{p_0}{m}T\right)+\chi_{-p_0}\left(-\frac{p_0}{m}T\right)\right]\otimes \ket{\psi(T)},
\end{equation}
with 
\begin{equation}
\ket{\psi(T)} = \sum_i \alpha_i e^{-\frac{i}{\hbar}E_iT\left(1- \frac{{p_0}^2}{2 m^2 c^2}\right)}\ket{\psi_i}.
\end{equation}
Therefore, as a result of the fact that both components of the superposition have the same value of $p^2$, the state continues to be separable, so tracing over the internal degrees of freedom does not yield an approximately diagonal reduced density matrix for the center of mass. That is, the decoherence effect reported in Ref.~\onlinecite{Pik:15a} is not present in this situation.

\begin{figure}[h]
\centering
\begin{pspicture}(6,5)
\psline[linewidth=.5pt](2.5,-.2)(2.5,5)
\psline[linewidth=.5pt](-1,0)(6,0)
\psline[linewidth=1pt](2.5,0)(.5,4)
\psline[linewidth=1pt](2.5,0)(4.5,4)
\psline[linewidth=.5pt](-1,3)(6,3)
\psdot[dotsize=5pt](2.5,0)
\psdot[dotsize=5pt](1,3)
\psdot[dotsize=5pt](4,3)
\rput(2.5,5.2){$x=0$}
\rput(6.6,0){$t=0$}
\rput(6.6,3){$t=T$}
\end{pspicture}
\caption{Trajectories of the wave packets in the $K$ frame.}
\label{fig1}
\end{figure}

Now consider the same physical system as described from the reference frame $K'$ which is moving with constant velocity $v$ with respect to $K$. Note that we are only making a passive transformation in which the system is not affected and only the reference frame is changed. Taking the coordinates in $K'$ so that its origin coincides with that of $K$, the coordinate transformation between $K$ and $K'$ is given by
\begin{eqnarray}
x' &=& (x-vt)\left(1+\frac{v^2}{2 c^2}\right), \\
t' &=& t\left(1+\frac{v^2}{2 c^2}\right)-\frac{vx}{c^2}.\label{tprime}
\end{eqnarray}
Note that, as above, we are keeping $c^{-2}$ terms because they may have an effect in the nonrelativistic regime. In $K'$, the description of the initial state is given by
\begin{equation}
\ket {\Psi' (0)} = \frac{1}{\sqrt{2}}\left[\chi'_{p_0-mv}(0)+\chi'_{-p_0-mv}( 0)\right]\otimes \ket{\psi(0)},
\end{equation}
which is still separable and has both wave packets centered at $x' =0$, one with momentum $p' = p_0-mv $ and the other with momentum $ p' = -p_0-mv$ (see Fig.~\ref{fig2}). Note that, since the initial state is localized around $x=0$ and, at that point, the hypersurfaces $t=0$ and $t'=0$ coincide, the transformation of the initial state from $K$ to $K'$ is quite simple and direct. Using the Hamiltonian (\ref{Hinitial}) it is easy to see that, at time $t' =T $, the state of the system (again, omitting a global phase) is given by
\begin{eqnarray}
\ket{\Psi'(T)} &=& \frac{1}{\sqrt{2}} \chi'_{p_0-mv}\left(\frac{p_0}{m}T-vT\right) \otimes \ket{\psi_+(T)}\nonumber\\
&+&\frac{1}{\sqrt{2}}\chi'_{-p_0-mv}\left(-\frac{p_0}{m}T-vT\right) \otimes \ket{\psi_-(T)}
\end{eqnarray}
where
\begin{equation}
\ket{\psi_{\pm}(T)} = e^{\pm\frac{i}{\hbar}p_0 v T} \sum_i \alpha_i e^{-\frac{i}{\hbar}E_iT\left(1- \frac{{(\pm p_0-mv)}^2}{2 m^2c^2}\right)}\ket{\psi_i}.
\end{equation}
Therefore, at time $T$ in the frame $K'$, the center of mass and the internal degrees of freedom are entangled, so the state is no longer separable. As a result, if one traces over the internal degrees of freedom, one finds, just as in the examples of Ref.~\onlinecite{Pik:15a}, an almost diagonal reduced density matrix for the center of mass. That is, one finds an effect which is completely analogous to the one reported in Ref.~\onlinecite{Pik:15a}.

\begin{figure}[h]
\centering
\begin{pspicture}(6,5)
\psline[linewidth=.5pt](2.5,-.2)(2.5,5)
\psline[linewidth=.5pt](-1,0)(6,0)
\psline[linewidth=1pt](2.5,0)(-.25,4)
\psline[linewidth=1pt](2.5,0)(3.75,4)
\psline[linewidth=.5pt](-1,3)(6,3)
\psdot[dotsize=5pt](2.5,0)
\psdot[dotsize=5pt](.45,3)
\psdot[dotsize=5pt](3.45,3)
\rput(2.5,5.2){$x'=0$}
\rput(6.6,0){$t'=0$}
\rput(6.6,3){$t'=T$}
\end{pspicture}
\caption{Trajectories of the wave packets in the $K'$ frame.}
\label{fig2}
\end{figure}

The first thing we note is that neither gravity nor acceleration is necessary for the decoherence effect to appear and that something as simple as a relative velocity between components of the state is sufficient (an observation already made in Ref.~\onlinecite{Pik:15a}). The point we want to make, though, is that for the same physical setting, we find the effect when analyzing things in one inertial frame but that the effect is simply absent when the situation is analyzed in the other. What we have here, then, is the inertial analog of the gravitational case in which the effect appears when analyzing the situation in the static frame, but disappears when considering it in a free-falling one. It thus seems that the effect is, in fact, frame dependent. Does this indicate a breakdown with the principles behind the theories of special and general relativity? Is decoherence frame dependent? What is really going on?

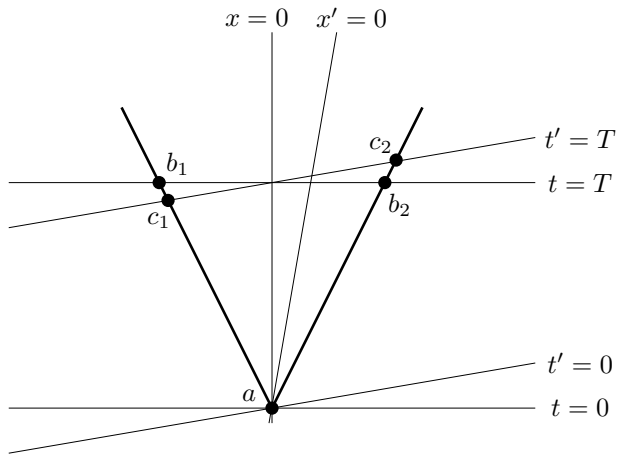
\begin{figure}[h]
\centering
\begin{pspicture}(6,6)
\psline[linewidth=.25pt](2.5,.3)(2.5,5.5)
\psline[linewidth=.25pt](2.46,.3)(3.36,5.5)
\psline[linewidth=.25pt](-1,.5)(6,.5)
\psline[linewidth=.25pt](-1,-.1)(6,1.1)
\psline[linewidth=.25pt](-1,3.5)(6,3.5)
\psline[linewidth=.25pt](-1,2.9)(6,4.1)
\psline[linewidth=1pt](2.5,.5)(.5,4.5)
\psline[linewidth=1pt](2.5,.5)(4.5,4.5)
\psdot[dotsize=5pt](2.5,.5)
\psdot[dotsize=5pt](1,3.5)
\psdot[dotsize=5pt](4,3.5)
\psdot[dotsize=5pt](1.118,3.263)
\psdot[dotsize=5pt](4.15,3.8)
\rput[dotsize=5pt](2.2,.675){$a$}
\rput[dotsize=5pt](1.25,3.75){$b_1$}
\rput[dotsize=5pt](4.2,3.2){$b_2$}
\rput[dotsize=5pt](1,3){$c_1$}
\rput[dotsize=5pt](3.95,4){$c_2$}
\rput(2.3,5.7){$x=0$}
\rput(3.56,5.7){$x'=0$}
\rput(6.6,.5){$t=0$}
\rput(6.6,3.5){$t=T$}
\rput(6.6,1.1){$t'=0$}
\rput(6.6,4.1){$t'=T$}
\end{pspicture}
\caption{Trajectories of the wave packets and constant-time hypersurfaces of $K$ and $K'$. Observers in $K$ compare proper times between segments $(a,b_1)$ and $(a,b_2)$, finding no difference; observers in $K'$ compare segments $(a,c_1)$ and $(a,c_2)$, finding a difference.}
\label{fig3}
\end{figure}

\begin{figure}[h]
\centering
\begin{pspicture}(6,5)
\pscurve[linewidth=.25pt](0.5,0)(.6,1.789)(.7,2.53)(.8,3.098)(.9,3.578)(1,4)
\pscurve[linewidth=1pt](2,0)(2.1,1.789)(2.2,2.53)(2.3,3.098)(2.4,3.578)(2.5,4)
\pscurve[linewidth=1pt](3.5,0)(3.6,1.789)(3.7,2.53)(3.8,3.098)(3.9,3.578)(4,4)
\psline[linewidth=.25pt](-0.5,2.760)(5,4.072)
\psline[linewidth=.25pt](-0.5,0)(5,0)
\psline[linewidth=.25pt](-0.5,2.9)(5,2.9)
\rput(6,0){$t=t'=0$}
\rput(6,3){$t=T$}
\rput(6,4.072){$t'=T$}
\rput(-0.5,2){Trajectory}
\rput(-0.5,1.6){of a}
\rput(-0.5,1.2){static}
\rput(-0.5,0.8){observer}
\psdot[dotsize=5pt](2,0)
\psdot[dotsize=5pt](3.5,0)
\psdot[dotsize=5pt](2.281,2.9)
\psdot[dotsize=5pt](3.781,2.9)
\psdot[dotsize=5pt](3.5,0)
\psdot[dotsize=5pt](2.38,3.42)
\psdot[dotsize=5pt](3.972,3.82)
\rput[dotsize=5pt](1.75,.2){$a_1$}
\rput[dotsize=5pt](2,2.6){$b_1$}
\rput[dotsize=5pt](2.15,3.65){$c_1$}
\rput[dotsize=5pt](3.8,.2){$a_2$}
\rput[dotsize=5pt](4,2.6){$b_2$}
\rput[dotsize=5pt](3.7,4){$c_2$}
\end{pspicture}
\caption{Free-falling observers compare proper times between segments $(a_1,b_1)$ and $(a_2,b_2)$, finding no difference; static observers compare segments $(a_1,c_1)$ and $(a_2,c_2)$, finding a difference.}
\label{figp}
\end{figure}
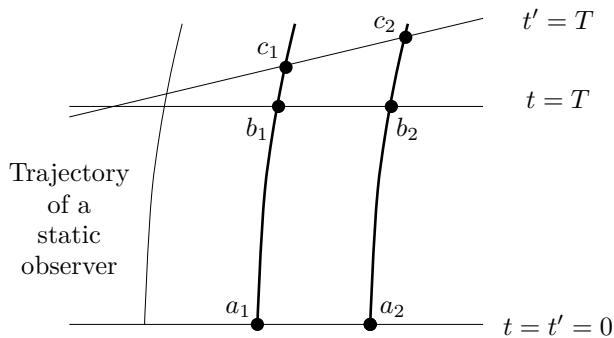

A key observation that helps clarify the seemingly contradictory conclusion regarding the presence of decoherence in $K'$, but not in $K$ (or its presence in a static frame, but not in a free-falling one), is that, in each frame, the initial and final states under consideration are tied to \emph{different} pairs of spatial hypersurfaces: $t=0$ and $t=T$ and $t'=0$ and $t'=T$, respectively (Fig.~\ref{fig3}). Therefore, $K$ and $K'$ and are not describing the same physical situation from two perspectives but two different situations involving the evolution of the system between two different pairs of events.\footnote{Note that the disagreement remains relevant in the nonrelativistic regime despite the fact that it is of order $c^{-2}$. As we explained above, that is because such time differences are multiplied by the mass, which involves a $c^{2}$ factor.} Something exactly analogous occurs in the case involving gravity, where one can either analyze the situation from the point of view of an accelerated frame (for instance, the frame associated with a static observer on the Earth's surface) or from the perspective of an inertial (\textit{i.e.}, free-falling) frame. In this case, the descriptions in both frames also correspond to different physical situations; the different worldline segments of the two wave packets involved, considered in each of the frames, are schematically shown in Fig.~\ref{figp}.

From the above discussion it is clear that the presence or absence of the decoherence effect reported in Ref.~\onlinecite{Pik:15a} crucially depends on the pair of hypersurfaces used to describe the system; the result arises when one considers a pair of hypersurfaces such that there is a difference in the elapsed proper times of the wave packets involved. This is the case for the frame $K'$, and for the static observers of the examples in Ref.~\onlinecite{Pik:15a}, but not for the frame $K$, or the free-fall observers. Furthermore, it seems clear that the reported effect is something fairly easy to achieve because, given a complex system and any background spacetime, one can always find a pair of hypersurfaces that will display the required difference in proper times that leads to the decoherence effect. The important question, then, is whether such an effect is physically substantial. In order to answer, in the next section we explore the conditions required to produce the effect, and we check if the claims of universality stated in Ref.~\onlinecite{Pik:15a} actually hold. 

\section{Is the effect physically meaningful?}
\label{Wit}

As we pointed out in the previous section, it is clear that the type of effect reported in Ref.~\onlinecite{Pik:15a} will be present whenever there is a difference of proper times between the initial and final events along the worldlines of the two wave packets making up the quantum state under consideration. However, it is crucial to recognize that, if one is given one of the worldlines in question and an initial and a final point on it, there is no principled or canonical way to select the corresponding initial and final points on the other worldline. That, of course, is a consequence of the fact that, in a general relativistic context, there are no preferred foliations of spacetime. This, together with the strong dependence of the effect on the exact segments of the worldlines that one uses to perform the calculations, leads us to question the applicability of the results reported in Ref.~\onlinecite{Pik:15a} outside the regime of specific experimental settings, purposely designed to measure interference. Consequently, in this section we explore the validity of claims in Ref.~\onlinecite{Pik:15a} regarding the generality and universal validity of their findings.

In order to clarify our discussion regarding the arbitrariness involved in the selection of the initial and final hypersurfaces, we note that for a generally covariant theory, like quantum field theory on a fixed curved spacetime, one is free to consider the evolution between arbitrary Cauchy hypersurfaces. In the case of free fields, this aspect is sometimes obscured by the fact that one usually relies on the Heisenberg formulation of the theory where operators, and, in particular, field operators, are spacetime dependent, while the quantum states are fixed (and unchanging). The situation becomes much clearer when interactions are included, and where the field evolution is again controlled by the free Hamiltonian $H_{\rm free}$, while the quantum states evolve according to the interaction Hamiltonian $H_{I}$. An essential aspect of a covariant interaction described by a local field theory is that, in the interaction picture, the change of the state resulting from an interaction can be described in terms of a Tomonaga-Schwinger evolution equation, which, in an infinitesimal form, is given by
\begin{equation}
i \delta \ket{\Psi (\Sigma)} = \mathcal{H}_I(x) \ket{\Psi (\Sigma)} \delta \Sigma(x),
\end{equation}
with $\mathcal{H}_I(x)$ a Hamiltonian interaction density constructed locally out of field operators. This equation gives the difference in the quantum state associated with the hypersurface $\Sigma'$ and that associated with the hypersurface $\Sigma$ when the former is obtained from the later by an infinitesimal deformation with four volume $\delta \Sigma(x)$ around the point $x$ in $\Sigma$.\footnote{We are ignoring the formal aspects regarding the fact that, strictly speaking, the interaction picture does not exist \cite{earman}.}
In this manner, the state associated with any Cauchy hypersurface can be obtained form the state on any other hypersurface using a finite composition of the infinitesimal version provided by the equation above. 

This formulation shows that such an evolution equation generates results that are clearly independent of the choice of coordinates and the interpolating foliation of spacetime. In turn, this reveals that, it would be impossible for any nontrivial entanglement of the quantum state, on any hypersurface, to depend on the coordinates or reference frame employed to describe the system and its evolution. Moreover, the above discussion makes it evident that the result completely depends on the specific hypersurface selected to be the final one, assuming the initial hypersurface is given. And the crucial point is that, in a general relativistic context, the choice of the specific hypersurfaces one uses to describe the system is completely arbitrary. The upshot of all this is that, any effect that so delicately depends on the choice of hypersurface, cannot be physically significant. 

One might argue that the choice of hypersurface is not as arbitrary as the preceding discussion suggests. That is because, intuitively, there seems to be a correspondence between different observers, and particular foliations of spacetime. Note, however, that this intuition arises from the notion of simultaneity usually considered in the contexts of Newtonian mechanics and its direct and naive extension to inertial frames in special relativity. However, this notion of simultaneity is generically not extensible to general relativity. In such a setting, any association of a given observer with a particular foliation is completely arbitrary. As a result, one cannot argue that any of the results reported in Ref.~\onlinecite{Pik:15a}, not explicitly related to a well-defined experiment, correspond to what any particular observer will perceive.

The point we want to make is that, in order to analyze the question of the presence or absence of interference effects between two localized components of a quantum superposition, one needs a precise description of where and how one would attempt to detect it. For instance, if the two wave packets are to be brought together, as is done in standard interference experiments, then, the detailed evolution of the state required for taking these two wave packets to a single position must also be taken into the account. Only then would one be able to decide whether interference would occur or not. Alternatively, if one is just told that, on a given hypersurface, the state is entangled, there is nothing one can conclude in terms of the absence or presence of interference for the system in question. 

The key observation is that, even if one knows that on a given hypersurface the state of the system is entangled, in general, further nontrivial evolution of the quantum state occurs when bringing to a common spacetime event the two components that were spatially separated. And, of course, this evolution can enhance or cancel the relevant phase differences. To avoid having to bring the two components of the state to a single spacetime event, one could perhaps consider some sort of nonlocal measurements. However, these nonlocal measurements are severely restricted because, if they were possible, one could use them for noncausal signaling \cite{Sorkin}. All this casts serious doubts on the significance of the arguments made in Ref.~\onlinecite{Pik:15a} about fundamental decoherence and, in particular, about their relevance regarding generic experimental situations. The main point, then, is that, without a detailed description of a specific experimental setup, one cannot say anything regarding the observability of interference patterns and, as a consequence, the claims about universal decoherence cannot be taken at face value.

\begin{figure}[h]
\centering
\begin{pspicture}(6,4)
\psline[linewidth=.5pt](2.5,-.2)(2.5,3.5)
\rput(6.6,3){$t=2T$}
\rput(6.6,0){$t=0$}
\rput(6.6,2.25){$t'=T$}
\rput(2.5,3.7){$x=0$}
\psline[linewidth=.25pt](-1,0.75)(6,2.25)
\psline[linewidth=.25pt](-1,0)(6,0)
\psline[linewidth=.25pt](-1,3)(6,3)
\psline[linewidth=1pt](2.5,0)(.5,4)
\psline[linewidth=1pt](2.5,0)(4.5,4)
\psdot[dotsize=5pt](2.5,0)
\psdot[dotsize=5pt](1,3)
\psdot[dotsize=5pt](4,3)
\psdot[dotsize=5pt](1.825,1.35)
\psdot[dotsize=5pt](3.37,1.675)
\end{pspicture}
\caption{The reported effect disappears if one first evolves the system to a hypersurface of constant $t'$ and then back into a hypersurface of constant $t$.}
\label{fig5}
\end{figure}
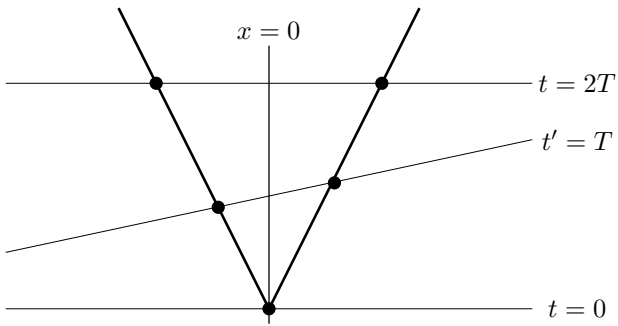

In order to illustrate the above discussion, 
we briefly come back to the simple example of Sec. \ref{exa}. There, even if the state of the system associated with the hypersurface $t'=T$ exhibits the kind of effect in question, it is clear that the system can be further evolved to a future hypersurface, say $t = 2T$, which is such that the corresponding quantum state once again is separable (see Fig.~\ref{fig5}). In fact, something completely analogous can be used to erase any nontrivial entanglement of the type considered in Ref.~\onlinecite{Pik:15a} which arises on the state on a given hypersurface: one only needs to exchange the locations of the two components for a suitable time to recover the full coherence of the initial state. Thus, in contrast with other types of decoherence which could be, in practice, irreversible, the present kind is always reversible by a simple procedure, further indicating that the claimed effect is not physically significant.

\begin{figure}[h]
\centering
\begin{pspicture}(6,4)
\psline[linewidth=.5pt](2.5,-.2)(2.5,3.5)
\psline[linewidth=1pt](1.25,0)(1.25,3.5)
\psline[linewidth=1pt](3.75,0)(3.75,3.5)
\rput(6.6,0){$t=0$}
\rput(2.5,3.75){$x=0$}
\rput(1.25,-0.35){$-x_0$}
\rput(3.75,-0.35){$x_0$}
\rput(6.8,2){$t=F(x)$}
\pscurve[linewidth=.5pt](-1,2)(1.25,1.5) (3.75,2.5)(6,2)
\psline[linewidth=.25pt](-1,0)(6,0)
\psdot[dotsize=5pt](1.25,0)
\psdot[dotsize=5pt](3.75,0)
\psdot[dotsize=5pt](1.25,1.5)
\psdot[dotsize=5pt](3.75,2.5)
\end{pspicture}
\caption{Spacetime diagram of the situation where the centers of mass of the two wave packets are at rest in the corresponding inertial frame and where the proper-time difference is only due to the particular form of the final hypersurface.}
\label{fig6}
\end{figure}
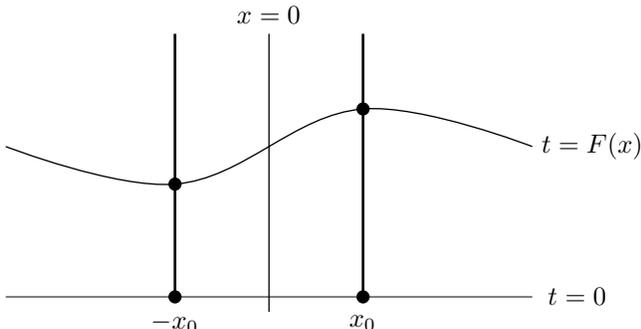

We may see how easy it is to produce the reported decoherence effect, and its triviality, in a still simpler case where there is no Newtonian gravitational potential and not even relative motion between the two localized components of the state. Consider a composite system that is at rest in a delocalized superposition corresponding to two wave packets located at $ x=x_0 $ and $ x=- x_0$. It is described by the initial state 
\begin{equation} \label{InStStat} 
\ket {\Psi (0)} = \frac{1}{\sqrt{2}}\left[\chi_{0}(x_0)+\chi_{0}(-x_0)\right]\otimes \ket{\psi(0)},
\end{equation}
with $\ket{\psi(0)}$, again, an arbitrary state for the internal degrees of freedom. This represents the state at hypersurface $t=0$ as described in a given inertial frame. Now let us consider the state associated with any hypersurface given by $t= F(x)$ such that $ F(-x_0) \neq F(x_0)$, as depicted in Fig.~\ref{fig6}. It easy to see that, in such a case, we would obtain essentially the same effect as in Ref.~\onlinecite{Pik:15a}. This clearly shows that whatever the significance of the result obtained, it cannot be claimed to reflect any novel aspect of physics intrinsically tied to gravitation, acceleration, or even relative motion of the wave packets.

The crucial point to note is that, in order to claim that a relevant decoherence is, in fact, taking place, \textit{i.e.}, that an effect which leads to an experimentally verifiable absence of interference is present, it is not enough to show that the state of the system in question, as described by some observer, leads to an almost diagonal reduced density matrix for the relevant degrees of freedom. As we showed above, this is easy to achieve by choosing appropriate hypersurfaces to describe the evolution of the system. What one needs to do in order to justifiably claim that one has found a substantial effect is to describe a concrete experiment where the interference disappears. Again we recall that in experiments that search for signatures of interesting quantum-mechanical interference patterns, one needs to bring the system in question to a common spacetime event where the detection is performed; only then the phase difference becomes relevant. This, of course, is in line with the fact that comparing phases associated with the quantum state of a system at different spacetime points is, on its own, essentially meaningless.

We believe that an important part of the misunderstanding regarding the decoherence claims in Ref.~\onlinecite{Pik:15a} arises from a lax use of the notion of decoherence and, in particular, from using the same term for two conceptually different matters. On the one hand, decoherence is used to describe a situation in which one takes the state of a composed system, as described from a given reference frame, and finds that, after tracing over some degrees of freedom that one is not interested in, the reduced density matrix characterizing the state for the rest of the system is almost diagonal. We argued above that this effect, which is frame dependent, is not really physically relevant. On the other hand, decoherence can be used to describe a situation in which, in a specific and well-defined experimental setting, the interference between different components of a superposition disappears. This second effect, which is frame independent, is the one that is substantial and physically significant. Our claim is that Ref.~\onlinecite{Pik:15a} only manages to show that the general scenarios considered lead to the first type of effect, but not the second. 

We conclude from all this that the claims in Ref.~\onlinecite{Pik:15a} regarding the generality and universality of the reported effect are invalid. In the next section we analyze if, in spite of all this, one can still claim that gravity is somehow responsible for the emergence of classicality.

\section{Can gravity account for the emergence of classicality?}
\label{sub}

In this section we finally scrutinize the strongest claim in Ref.~\onlinecite{Pik:15a} which holds that, through the decoherence effect reported, gravity accounts for the emergence of classicality. This is, of course, an extraordinary assertion with far reaching consequences. Unfortunately, a detailed inspection of the claim quickly shows that it is not warranted. In particular, below we argue that, at least in the scenarios discussed in Ref.~\onlinecite{Pik:15a}, one cannot say that gravitation is the agent responsible for the cited effect. In a nutshell, that is because none of the situations discussed involve the curvature of spacetime, which, according to our best description of gravity provided by general relativity, is the only manifestation of truly gravitational phenomena. Moreover, we offer a discussion of the errors underlying the widespread belief that decoherence by itself is enough to account for the quantum-to-classical transition. As we argue below in detail, the most common mistake in this regard arises from assigning an incorrect physical meaning to the reduced density matrix of a quantum subsystem.

\subsection{Gravity and decoherence}

As we pointed out above, in none of the scenarios considered in Ref.~\onlinecite{Pik:15a} does spacetime curvature play a relevant role. In fact, all frames where gravity appears in the authors' treatment are flat spacetimes, described in terms of noninertial coordinates. Therefore, one can always analyze these situations from local inertial frames, implemented through Riemann normal coordinates, in which gravity is absent. Moreover, it follows from the equivalence principle that a consistent treatment of these situations should lead to the same result whether it is carried out using one frame or the other.

From the previous observations it follows that the results in Ref.~\onlinecite{Pik:15a} are to be determined not by the presence of gravity, which as we have indicated can be removed from the picture, but by the effects that arise due to the external device used to control the motion of the system and the manner in which the interference detection is carried out. That is because, in order to carry out a particular experiment, an external device must be used in order to guide the system. However, by passing to a description of the situation as viewed from a free-falling frame, the force generated by the device is the only element present that could play a role in the appearance of any relevant effect. Of course, it is possible that the interaction of the system with such a device leads to some robust kind of decoherence, but it is clear that the result would depend on the exact nature of the interaction between the system and the external device, which has nothing to do with gravity.

It is clear, then, that, while some kind of decoherence might happen in specific experimental situations, gravity, as envisioned in Ref.~\onlinecite{Pik:15a}, is never the agent that causes the effect. In fact, only by clinging to a Newtonian point of view it would make sense to assume that, in situations where spacetime curvature is irrelevant, there are gravitational effects. On the other hand, we must acknowledge that, in highly nontrivial spacetimes, particularly in spacetimes where more than one geodesic joins two events, gravity might produce exotic effects. One interesting example is provided by the spacetime associated with a rotating planet or star and a quantum system in a superposition of a corotating and a counterrotating orbit. In such a case, each trajectory has a different proper time between successive encounters. The interesting feature, in this case, is that the two components are in free fall, and hence no external device is required to keep them orbiting and meeting at different spacetime points (as it would be needed to carry out an interference experiment).

\subsection{Decoherence and the emergence of classicality}

Finally, on top of all this, there is the question of whether decoherence is sufficient to explain the emergence of classicality. In other words, we now analyze if the mere presence of an almost-diagonal reduced density matrix is enough to explain the quantum-to-classical transition. The standard justification of this idea is that, since the reduced density matrix in question has the same form as a matrix representing a statistical mixture, then, \emph{for all practical purposes} (or FAPP, as coined by Bell in Ref.~\onlinecite{Bell}), the center of mass behaves as such a mixture and does not display a ``strange'' quantum behavior. Unfortunately, the answer to the question at the beginning of the paragraph is negative (see \emph{e.g.}, \cite{deco1,deco2}). 

To start off, we already saw that whether a reduced density matrix of the type considered in Ref.~\onlinecite{despagnat} is almost diagonal or not is highly dependent on the hypersurface on which one describes the state. That is, the reduced density matrix can be diagonal in one frame but not in the other. From this one must conclude that decoherence of this type does not lead to classicalization; otherwise one would have to accept that classicality depends on the frame employed to describe the system.

At any rate, the real reason behind the inability of decoherence to explain classicality is that, even if it is the case that the reduced density matrix for the center of mass, after tracing over the internal degrees of freedom, has the same form as that which represents a statistical mixture, in which the center of mass is in \emph{either} one or the other of the locations of the superposition, it does not follow that the \emph{physical} situation is identical. In fact, the physical situation is extremely different since, by taking the trace, or by the mere act of ignoring the internal degrees of freedom for that matter, clearly nothing physical happens to the system, and thus, the physical state of the center of mass continues to be as entangled and delocalized as it was before.

The central point is that one should be careful to distinguish a \emph{proper} mixture, which represents an actual ensemble of systems, each of which has been prepared to be in a different but well-defined state, from an \emph{improper} mixture, which represents the partial description, as provided by the reduced density matrix, of a subsystem which is part of a larger system that, as a whole, is in a pure state (see, \textit{e.g.}, Ref.~\onlinecite{despagnat}).

A simple example that makes this point evident is provided by the following Einstein-Podolsky-Rosen (EPR) setup. Consider the decay of a spin $0$ particle into two spin $1/2$ particles, and take the $x$ axis as the direction of the decay (the momenta of the particles are $\vec{P}= \pm P \hat{x}$ with $\hat{x}$ the unit vector in the $x$ direction). After the decay, the two-particle state can be characterized in terms of the $z$-polarization states of the two individual-particle Hilbert spaces. The angular momentum conservation implies that the state must be
\begin{equation}
|\phi \rangle= \frac{1}{\sqrt2} \left( |+ \rangle^{(1)}_{z} |- \rangle^{(2)}_{z} + |-\rangle^{(1)}_{z} |+ \rangle^{(2)}_{z} \right),
\end{equation}
where we denote by $|\pm \rangle^{(i)}_{z}$ the $i$th particle spin state in the $\pm z$ direction. It is easy to see that such a state is invariant under rotations around the $x$ axis because it is an eigenstate of angular momentum along that axis (with eigenvalue $0$). As always, the density matrix for the system is given by $\rho= |\phi\rangle\langle\phi|$. 

Now, assume that we decide to ignore one of the particles, say particle $1$, and thus we regard it as an environment for the system of interest, namely, particle $2$. The reduced density matrix for particle $2$, obtained by tracing over the degrees of freedom of particle $1$, is then
\begin{equation}
\rho^{(2)} = \frac{1}{ 2}\left( |+ \rangle^{(2)}_{z} \langle+|^{(2)}_{z}+|- \rangle^{(2)}_{z} \langle-|^{(2)}_{z} \right) ,
\end{equation}
which clearly is diagonal. Therefore, we have obtained a fully decohered density matrix. Does this mean that the particle must be considered as having a definite value of either $+1/2 $ or $-1/2$ for its spin along the $z$ axis? Of course not. First of all, it is clear given Aspect's experiments confirming the violation of Bell's inequalities \cite{Aspect1,Aspect2,Aspect3} that one cannot assume that such a particle has a definite (even if unknown) value for its spin before a measurement takes place. That is, decoherence is not equivalent to classicality. Furthermore, the fact that the state $|\phi\rangle$ is symmetric with respect to rotations around the $x$ axis implies that we could have written the density matrix as
\begin{equation}\label{DM}
\rho^{(2)} = \frac{1}{ 2}\left( |+ \rangle^{(2)}_{y} \langle+|^{(2)}_{y}+|-
\rangle^{(2)}_{y} \langle-|^{(2)}_{y} \right)
\end{equation}
leading, this time, to the conclusion that the particle spin is either $+1/2 $ or $-1/2$ along the $y$ axis and not the $x$ axis.

Another simple example that clearly illustrates the fact that proper and improper mixture have extremely different physical meanings is given by a particle with spin living in a plane with state given by
\begin{equation}\label{CMS}
|\varphi \rangle= \frac{1}{\sqrt2} \left( \chi_{\vec{0}}(x_0,0) |+ \rangle_{y} + \chi_{\vec{0}}(-x_0,0) |- \rangle_{y} \right) ,
\end{equation}
with $\chi_{\vec{p}}(x,y)$ a wave packet centered in $(x,y)$ with momentum $\vec{p}$. This state represents a particle in a superposition of being centered around $(x_0,0)$ with positive spin along $y$, and centered around $(-x_0,0)$ with negative spin along $y$. It is clear that such a state is symmetric under rotations by $180^\circ$ along the $z$ axis. Now, in analogy with what is done in Ref.~\onlinecite{Pik:15a}, suppose that we decide to ignore the internal degrees of freedom of the system, and we thus trace over the spin. Of course, the resulting density matrix is diagonal, but, does this mean that the particle is now located in either $(x_0,0)$ or $(-x_0,0)$? Again, the answer is negative: it is clear that the simple act of deciding to ignore the spin of the particle does not, in any way, change the fact that the particle is in a superposition of position. Evidently, the same considerations apply to all the systems considered in Ref.~\onlinecite{Pik:15a}. That is, the fact that one decides to ignore the internal degrees of freedom does not change the fact that the center of mass is in a superposition of positions. Notice that the simplicity of the examples considered here does not mean that the conceptual conclusions extracted are different when the analogous considerations are done in more complex systems. That is, even for complex systems, by taking a partial trace over some degrees of freedom, the approximate (or even exact) diagonal form of the resulting reduced density matrix does not indicate that the system can be considered as a single element of a classical ensemble (not even if by the usage of the word classical we mean that each element of the ensemble has a well-defined value of the quantity of interest, which in our case is the position).

In spite of all this, one may still claim that the fact that decoherence produces an approximately diagonal density matrix can be taken as indicating that, even though proper and improper mixtures are physically different, they behave in the same way as far as empirical predictions are concerned, \textit{i.e.}, FAPP. Several comments are in order. First and foremost, extreme care is required when using the FAPP sentence since, in some situations, it might be impossible to manipulate the degrees of freedom one has traced over, at least with current or even conceivable technology. However, that is evidently not the case in the EPR example we mentioned above where a subsequent study of the correlations between the spin degrees of freedom of the two particles reveals, in an empirical way, that, despite the diagonal form of the density matrix for particle $2$, given in Eq.~(\ref{DM}), we always have an improper mixture. Moreover, the situations considered in Ref.~\onlinecite{Pik:15a} are of this type, as it can be elucidated from our discussions regarding the simple ways in which the interference between two spatially separated components of the system can be restored (see Fig.~\ref{fig5} and the discussion around it).
 
In addition, it is crucial to recognize that, in order to defend the claim that FAPP a system described by an improper mixture behaves as one described by a proper one (as it would be needed, in addition to decoherence, in order to truly explain the emergence of the classical regime), one has to assume, often inadvertently, that when one measures an observable, one does not find a superposition but one of the eigenvalues of the measured observable with probabilities given by Born's rule. Let us see this more explicitly. For once, it should be clear that, without any rule connecting the formalism with observations, quantum theory cannot make predictions. Next, note that one wants to explain, say in the second example described by Eq.~(\ref{CMS}), why, even though the center of mass is in a superposition, when one observes it, one finds a well-defined value for the center of mass position. The explanation offered by decoherence is that, since the reduced density matrix has the same form as that corresponding to proper mixture, then it behaves as such FAPP. However, note that the diagonal elements of a proper mixture do represent \emph{probabilities}, and thus if one wants to interpret the diagonal elements of an improper mixture in precisely the same way, \textit{i.e.}, as probabilities, one must invoke Born's rule. In other words, to argue that decoherence can explain why one finds a definite result when measuring position, one has to presuppose the Born rule. As a result, and contrary to what is commonly argued, to really \emph{explain} the classical behavior using decoherence, one necessarily needs to go beyond quantum mechanics and adopt, from the beginning, something like the Dirac \cite{Dirac} or von Neumann \cite{von} interpretations of the quantum theory, which indicate that the evolution of the quantum state during measurements departs from the purely unitary evolution of the Schr\"odinger equation.

Regarding these issues, in Ref.~\cite[p. 4]{Pik:15b} Pikovski \textit{et al.} state the following:
\begin{quotation}
Our results do not hinge on an alleged physical difference between ``proper'' and ``improper mixtures,'' which lies outside of quantum theory and the scope of our work.
\end{quotation}
Various comments are in order regarding this statement. First, it is hard to understand why they believe that this issue lies outside of quantum theory. In fact, if one is interested in the emergence of classicality, it seems paramount to be careful in separating an alive and a dead cat from one that is in a quantum superposition of the two states. In fact, when considering the standard Schr\"odinger's cat setup, after tracing over the state of the proverbial atom (the nucleus of which is characterized by the two subspaces of the Hilbert space corresponding to a decayed and an undecayed state), we find an improper mixture, while, if we consider, instead, an ensemble of cats, half of which are definitely alive and with the other half being definitely dead, we would be dealing with a proper mixture. The distinction is essential in any discussion aimed to explain the emergence of classicality. As clearly stated by Bell: \emph{and} is not the same as \emph{or} \cite{Bell}. Thus, it must be clear from the examples discussed above that the physical difference between proper and improper mixtures is, in fact, real and easy to grasp, and that it is crucial to evaluate the ability of decoherence to account for the emergence of classicality. Finally, in our view, the fact that the discussion of these extremely relevant conceptual matters is declared by the authors of Ref.~\onlinecite{Pik:15a} to lie outside of the scope of their work illustrates their unwillingness to critically assess the conceptual significance of their analysis.

We conclude from all this that the extraordinary claim in Ref.~\onlinecite{Pik:15a} regarding the ability of gravity to account for the emergence of classicality is not valid.

\section{Conclusions}
\label{Con}

With the aid of some simple examples not involving gravitation or acceleration, we examined in detail some of the controversies surrounding the claim in Ref.~\onlinecite{Pik:15a} that gravitation produces a universal type of decoherence for systems with internal degrees of freedom, which might account for their quantum-to-classical transition. In this regard, we first considered an apparent conflict between the reported effect and the equivalence principle, which suggests that the effect is frame dependent. We clarified the issue by pointing out that the apparent conflict arises from not recognizing that the descriptions of the scenarios considered, as given by different reference frames, do not correspond to alternative descriptions of the same physical situation but to descriptions of different physical settings. Then, in order to avoid this type of confusion, we underscored the fact that the effect is highly dependent on the initial and final hypersurfaces under consideration. We also emphasized the arbitrariness involved in choosing such hypersurfaces.

Next, we noted that one of the main problems with the conclusions reached in Ref.~\onlinecite{Pik:15a} is associated with a failure to note that, in order to discuss the interference between two components of a delocalized quantum state, one must be very precise in describing the specific experimental situation. As a result, we questioned the applicability of the reported calculations for generic scenarios and we stressed the inviability of claims in Ref.~\onlinecite{Pik:15a} regarding the universality of the effect.

Finally, we analyzed in detail the strongest claim in Ref.~\onlinecite{Pik:15a} regarding the capacity of gravity to explain the emergence of classicality. We found such an assertion wanting on, at least, two fronts. First, given the fact that the curvature of spacetime is never essential in the scenarios considered, we pointed out that gravity is in fact not responsible for the cited effect. Second, we explained the shortcomings of the pervasive belief that decoherence, by itself, can explain the quantum-to-classical transition. As a result from all of the above, we reach the conclusion that the claims in Ref.~\onlinecite{Pik:15a} regarding the universality of the cited effect, and its capacity to account for the emergence of classicality, are invalid.

\section*{Acknowledgments}
We acknowledge partial financial support from UNAM-DGAPA-PAPIIT Projects No. IA101116 (Y. B.), IA400114 (E. O.), IG100316 (E. O. and D. S.), and CONACyT Project No. 220738 (D. S.).

\end{document}